\documentclass[preprint,12pt]{elsarticle}

\usepackage[normalem]{ulem}  

\usepackage{amssymb}
\usepackage{amsmath}
\usepackage{braket}

\journal{Nuclear Physics B}

\begin{document}

\begin{frontmatter}



\title{Structure of near-threshold resonances \\with new interpretation scheme of complex compositeness}


\author{Tomona Kinugawa and Tetsuo Hyodo} 

\affiliation{organization={Department of Physics, Tokyo Metropolitan University},
            addressline={1-1 Minami-Osawa}, 
            city={Hachioji},
            postcode={192-0397}, 
            state={Tokyo},
            country={Japan}}

\begin{abstract}
The nature of near-threshold resonances is quantitatively studied with a new interpretation scheme using the complex compositeness. A difficulty was known in the understanding of the internal structure of unstable resonances because their complex compositeness is not an interpretable measure. To overcome this problem, we develop a new interpretation scheme respecting the ambiguous aspects of the identification of the internal structure of resonances. We then apply the interpretation scheme to the near-threshold resonances slightly above the threshold, described by the effective range expansion. With the new interpretation scheme, we show that near-threshold resonances are dominated by the non-molecular component. Namely, even in the near-threshold region, the nature of resonances is sharply contrasted with bound states whose internal structure is usually molecular dominant.
\end{abstract}




\end{frontmatter}



\section{Introduction}
\label{sec:intro}

In hadron physics, many experiments have reported candidates of exotic hadrons as near-threshold quasi-bound states~\cite{Hosaka:2016pey,Brambilla:2019esw,Chen:2022asf}. For example, a genuine exotic tetraquark state $T_{cc}$ is observed near the $D^{0}D^{*+}$ threshold as a meson with charm $C = +2$~\cite{LHCb:2021auc,LHCb:2021vvq}. In the $C = 0$ sector, $X(3872)$ is a well known near-threshold exotic hadron below the $D^{0}\bar{D}^{*0}$ threshold~\cite{Belle:2003nnu}. One of the possible internal structures of exotic hadrons is the hadronic molecule, a loosely bound state of hadrons~\cite{Guo:2017jvc}. When the wavefunction of an exotic hadron is dominated by the hadronic molecule component, we call the hadron molecular dominant.

To understand the nature of shallow bound states, low-energy universality plays an important role. When a sufficiently shallow bound state is formed below the threshold, the scattering length $a_{0}$ becomes large and any quantities in the system are scaled by $a_{0}$~\cite{Braaten:2004rn,Naidon:2016dpf}. This low-energy universality induces the threshold energy rule~\cite{PTPS.E68.464}, which asserts that the internal structure of shallow bound states are in general molecular dominant, regardless of the microscopic origin of the state~\cite{Hyodo:2014bda,Hanhart:2014ssa,Sazdjian:2022kaf,Lebed:2022vks,Hanhart:2022qxq,Kinugawa:2023fbf}. In contrast to the well-established results on shallow bound states below the threshold, near-threshold resonances above the threshold has not been studied very much. However, a recent analysis~\cite{Sarti:2023wlg} suggests the existence of such a near-threshold state $\Xi(1620)$ whose excitation energy is $+3.68$ MeV measured from the $K^{-}\Lambda$ threshold. Therefore, to elucidate the nature of exotic hadrons, it is desirable to study not only the internal structure of shallow bound states but also that of near-threshold resonances. One of the purposes of this study is to seek for the universal nature of near-threshold resonances.

To quantitatively study the molecular nature of exotic hadrons, the compositeness has been utilized. The compositeness $X$ is defined as the fraction of the hadronic molecular component $\ket{\rm molecule}$ in the wavefunction $\ket{\Psi}$~\cite{Weinberg:1965zz,Baru:2003qq,Hyodo:2011qc,Aceti:2012dd,Hyodo:2013iga}:
\begin{align}
\ket{\Psi} = \sqrt{X}\ket{\rm molecule} + \sqrt{Z}\ket{\rm non\ molecule}.
\end{align}
Here $Z = 1 - X$ is called the elementarity, the fraction of the non-molecular component $\ket{\rm non\ molecule}$. For stable bound states, the compositeness and elementarity serve as the useful measures to quantify the internal structure, because $X$ ($Z$) can be interpreted as a probability of finding the molecular (non-molecular) component. For example, in Refs.~\cite{Weinberg:1965zz,Kamiya:2017hni,Li:2021cue,Song:2022yvz,Albaladejo:2022sux,Kinugawa:2022fzn}, the deuteron is concluded as the $pn$ molecule state because its compositeness is close to unity. In fact, the compositeness of shallow bound states is usually $X\sim 1$ (i.e., molecular dominant) from the low-energy universality, as mentioned above~\cite{Hyodo:2014bda,Hanhart:2014ssa,Sazdjian:2022kaf,Lebed:2022vks,Hanhart:2022qxq,Kinugawa:2023fbf}.

However, almost all exotic hadrons are unstable resonances which decay by the strong interaction. Because the usual norm of an unstable state of the unstable state $\braket{\Psi|\Psi}$ diverges, the Gamow vector $\bra{\tilde{\Psi}}\neq \ket{\Psi}^{\dag}$ has to be introduced for the normalization $\braket{\tilde{\Psi}|\Psi} = 1$~\cite{Berggren:1968zz}. In this case, the compositeness $X$ and elementarity $Z$ are given by the complex values by definition~\cite{Kamiya:2016oao}:
\begin{align}
X &= \braket{\tilde{\Psi}|{\rm molecule}}\braket{{\rm molecule}|\Psi} \in \mathbb{C}, \quad Z = 1 - X \in \mathbb{C}.
\end{align}
The complex compositeness of unstable resonances cannot be regarded as the probability, and the interpretation of the internal structure of exotic hadrons is not straightforward. In this context, we need to develop a reasonable interpretation scheme of the complex compositeness and elementarity.

In this work, we discuss two issues in order to understand the nature of near-threshold resonances. We first propose a new interpretation scheme for unstable resonances with the complex compositeness in Sec.~\ref{sec:interpretation}. In Sec.~\ref{sec:near-th}, we then apply our interpretation scheme to near-threshold resonances described with the effective range expansion. The last section (Sec.~\ref{sec:summary}) is devoted to summary. The details of this work can be found in Ref.~\cite{Kinugawa:2024kwb}

\section{Interpretation scheme for resonances}
\label{sec:interpretation}

\begin{figure}[t]
\centering
\includegraphics[width=1.0\textwidth]{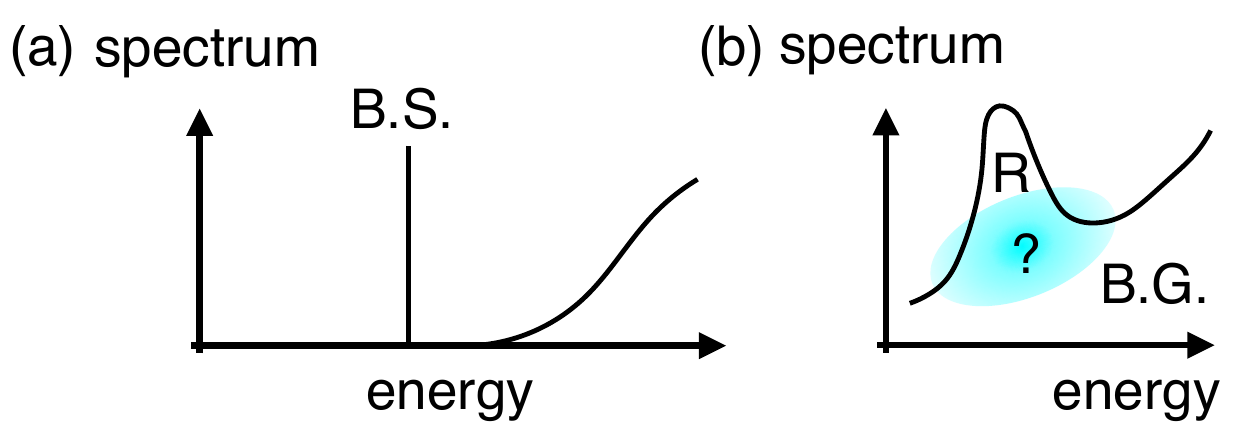}
\caption{
Schematic illustration of the transition spectrum with a bound state [B.S. in panel (a)] and that with a resonance [R in panel (b)]. It is uncertain whether the shaded components represent the resonance contribution or the background (B.G.) contribution.}
\label{fig:un}
\end{figure}

To develop a reasonable interpretation scheme of resonances, we refer to Ref.~\cite{Berggren:1970wto} where the author discusses the extraction of the resonance contribution from the transition process in nuclear reactions. If the intermediate state is a stable bound state, we can uniquely identify the bound state contribution from the spectrum, because the bound state appears as an isolated delta function without  width, which is separated from the continuum [Fig.~\ref{fig:un} (a)]. Mathematically, the bound state contribution is extracted by inserting the projection operator $P_{B}=\ket{\Psi}\bra{\Psi}$. In contrast, unstable resonance is embedded in the continuum spectrum with the finite width. As a consequence, uncertainty arises in the identification of the resonance contribution, because the resonance energy is not unique due to the finite lifetime and the separation from the background under the resonance contribution is ambiguous [Fig.~\ref{fig:un} (b)]. 
From this viewpoint, the author in Ref.~\cite{Berggren:1970wto} considers that the uncertainty is related to the complex nature of the resonance contribution extracted by the operator $P_{R}=\ket{\Psi}\bra{\tilde{\Psi}}$, and suggests to classify the outcome of an experiment into
\begin{itemize}
\item[(I)] certainly identified as the resonance;
\item[(II)] certainly identified as not the resonance;
\item[(III)] uncertain whether the resonance or not.
\end{itemize}
Namely, the identification of the intermediate states is not clear-cut when unstable resonances are involved. A prescription to relate the complex matrix element to the probabilities $a_{n}, b_{n}, c_{n}$ corresponding to (I), (II), (III) is suggested in Ref.~\cite{Berggren:1970wto}. We note that the bound state is classified only into (I) and (II) because the category (III) reflects the uncertainty of resonances.

We apply this idea to the extraction of the molecular component of a resonance from its wavefunction. For bound states, the internal structure is unambiguously identified as either molecular or non-molecular component, and the real compositeness $X$ can be interpreted as a probability. In contrast, based on the discussion in Ref.~\cite{Berggren:1970wto}, it is natural to consider that there are some ambiguities in the identification of the internal structure of resonances due to their unstable nature, which is reflected in the complex compositeness $X$. We therefore introduce a new category ``uncertain identification'' in addition to the molecular and non-molecular identifications of the internal structure of resonances;
\begin{itemize}
\item[(i)] certainly identified as the molecular component;
\item[(ii)] certainly identified as the non-molecular component;
\item[(iii)] uncertain whether the molecular component or not.
\end{itemize}

Let us recall the definition of the compositeness $X$ and elementarity $Z$ of bound states. $X$ and $Z$ are defined as the probability of finding the molecular component and non-molecular one, respectively. From the above argument, we introduce the following three probabilities $\mathcal{X,Z,Y}$ which characterize the identifications of (i), (ii), (iii), respectively;
\begin{itemize}
\item[$\mathcal{X}$]: probability of certain identification of the molecular component;
\item[$\mathcal{Z}$]: probability of certain identification of the non-molecular component;
\item[$\mathcal{Y}$]: probability of uncertain identification.
\end{itemize}
We regard $\mathcal{X}$ ($\mathcal{Z}$) as the compositeness (elementarity) of resonances instead of complex $X$ and $Z$. Because the category (iii) is introduced to represent the ambiguities in the identification of resonances, corresponding probability $\mathcal{Y}$ also reflects the uncertain nature of resonances. Therefore, $\mathcal{Y}$ should vanish for bound state without any uncertainties. 

In order to establish a reasonable interpretation scheme for resonances with $\mathcal{X,Y,Z}$, these quantities should naturally extend the concepts of the compositeness and elementarity of bound states. In other words, $\mathcal{X,Y,Z}$ reduce to the bound state case as
\begin{align}
\mathcal{X} \to X,\quad \mathcal{Z} \to Z, \quad \mathcal{Y} \to 0.
\end{align}
To satisfy this condition, we relate $\mathcal{X,Y,Z}$ with complex $X,Z$ as follows:
\begin{align}
\mathcal{X}+\alpha \mathcal{Y} &= |X|, 
\label{eq:XY}\\
\mathcal{Z}+\alpha \mathcal{Y} &= |Z|, 
\label{eq:YZ}
\end{align}
where $\alpha$ is an arbitrary positive parameter which controls the degree of ambiguity in resonances. The explicit value of $\alpha$ will be determined in the next section. For more detailed discussion on the parameter $\alpha$ in Eqs.~\eqref{eq:XY} and \eqref{eq:YZ}, please refer to the full paper~\cite{Kinugawa:2024kwb}.

For the probabilistic interpretation, $\mathcal{X,Y,Z}$ should be normalized as $\mathcal{X} +\mathcal{Y} + \mathcal{Z} = 1$. From Eqs.~\eqref{eq:XY}, \eqref{eq:YZ} and this normalized condition, we obtain the expressions of $\mathcal{X,Y,Z}$ using complex $X$ and $Z$:
\begin{align}
\mathcal{X}&=\frac{(\alpha-1)|X|-\alpha|Z|+\alpha}{2\alpha-1}, \label{eq:calX} \\
\mathcal{Y}&=\frac{|X|+|Z|-1}{2\alpha-1}, \label{eq:calY} \\
\mathcal{Z}&=\frac{(\alpha-1)|Z|-\alpha|X|+\alpha}{2\alpha-1}. \label{eq:calZ}
\end{align} 
To be regarded as probabilities, $\mathcal{X,Y,Z}$ should also be positive. Because of the triangle inequality, the numerator of $\mathcal{Y}$ is always positive. Thus, to ensure a positive value of $\mathcal{Y}$, we should choose $\alpha > 1/2$. Even in this parameter region, however, $\mathcal{X}$ and $\mathcal{Z}$ can be negative. For example, if $\mathcal{Y}$ is large enough, either $\mathcal{X}$ or $\mathcal{Z}$ becomes negative to satisfy $\mathcal{X} +\mathcal{Y} + \mathcal{Z} = 1$. In this case, we regard such a state with negative $\mathcal{X}$ or $\mathcal{Z}$ as unphysical, because large $\mathcal{Y}$ indicates large ambiguity due to the unstable nature. From this consideration, we introduce a new category ``non-interpretable'' in the interpretation scheme to classify the unphysical states. In this way, unphysical (non-interpretable) states are automatically distinguished from physical (interpretable) states by the values of $\mathcal{X,Y,Z}$ in this classification scheme.

\begin{figure}[t]
\centering
\includegraphics[width=0.75\textwidth]{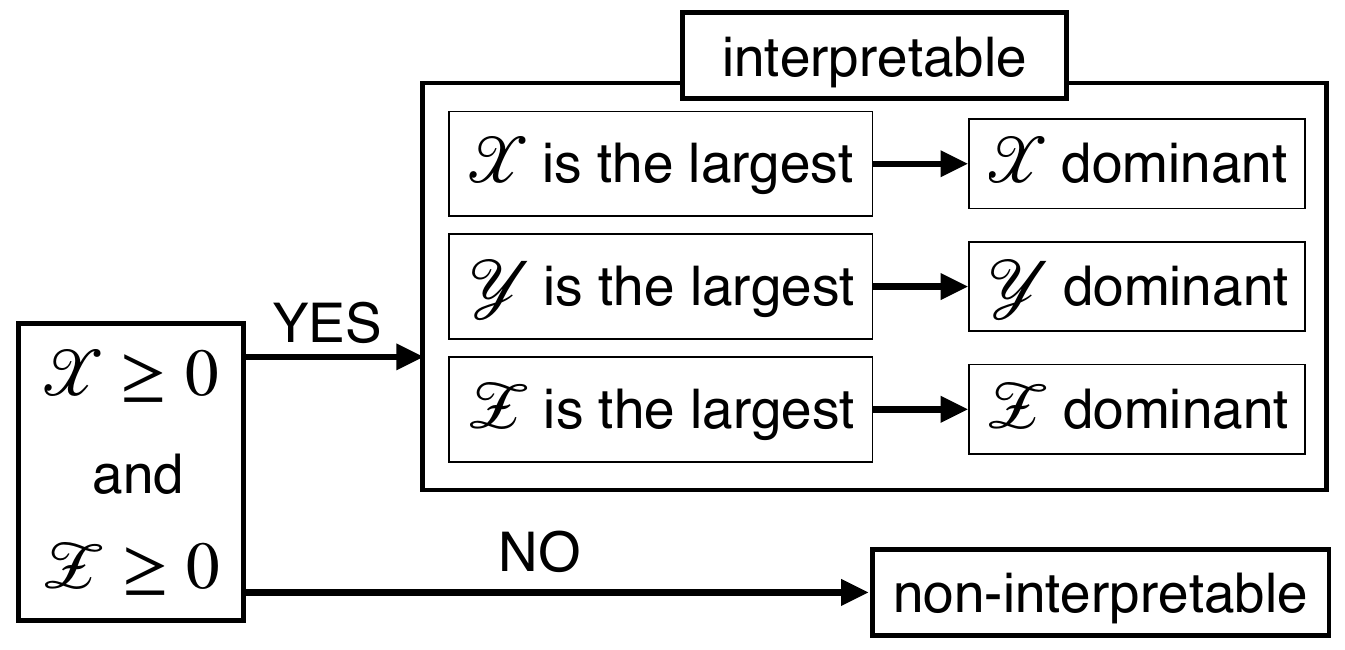}
\caption{The classification scheme of the internal structure of resonances 
with $\mathcal{X,Y,Z}$.}
\label{fig:interpretation}
\end{figure}

We summarize our interpretation scheme for resonances in Fig.~\ref{fig:interpretation}. If $\mathcal{X}$ or $\mathcal{Z}$ is negative, the resonance is non-interpretable. If $\mathcal{X,Y,Z}$ are all positive, the internal structure of the resonance is interpreted as being dominated by $\mathcal{X}$, $\mathcal{Y}$, or $\mathcal{Z}$, depending on which has the highest probability.

\section{Internal structure of near-threshold states}
\label{sec:near-th}

In this section, we discuss the internal structure of near-threshold resonances. For this purpose, we calculate the compositeness of the near-threshold resonances with the effective range expansion (ERE) and apply our interpretation scheme to these results. The ERE gives a general expansion of the scattering amplitude $f(k)$ truncated at the $k^{2}$ order, expressed by the scattering length $a_{0}$ and effective range $r_{e}$~\cite{Hyodo:2013iga}: 
\begin{align}
f(k)&=\left[-\frac{1}{a_{0}}+\frac{r_{e}}{2}k^{2}-ik\right]^{-1}.
\label{eq:f-ERE}
\end{align}
In the ERE, the eigenmomenta are given by the pole of the scattering amplitude $1/f(k) = 0$ as
\begin{align}
k^{\pm}&=\frac{i}{r_{e}}\pm\frac{1}{r_{e}}\sqrt{\frac{2r_{e}}{a_{0}}-1+i0^{+}}.
\label{eq:k-pm}
\end{align}
Here $k^{+}$ and $k^{-}$ are determined only by $a_{0}$ and $r_{e}$. Conversely, $a_{0}$ and $r_{e}$ can be expressed by $k^{\pm}$:
\begin{align}
a_{0}&=-\frac{k^{+}+k^{-}}{ik^{+}k^{-}}, 
\label{eq:a-ERE}\\
r_{e}&=\frac{2i}{k^{+}+k^{-}}.
\label{eq:r-ERE}
\end{align}

\begin{figure}[t]
\centering
\includegraphics[width=0.75\textwidth]{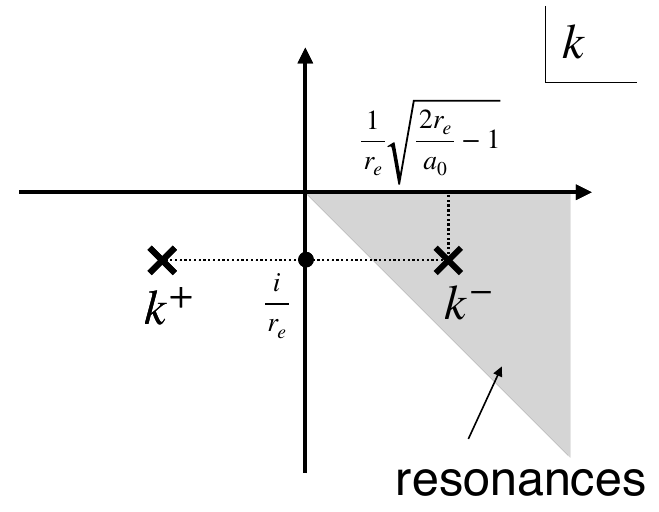}
\caption{Schematic illustration of the position of resonance pole $k^{-}$ and that of anti-resonance $k^{+}$ in the complex momentum $k$ plane. The resonance pole $k^{-}$ should be in the shaded region $0<-\theta_{k}\leq \pi/4$.}
\label{fig:resonance}
\end{figure}

The expressions of the eigenmomenta $k^{\pm}$~\eqref{eq:k-pm} correspond to bound states, virtual states, and resonances in the ERE by choosing suitable $a_{0}$ and $r_{e}$~\cite{Hyodo:2013iga}. To focus on resonances which appear in the $0 < -\theta_{k} \leq \pi/4$ region in the complex momentum plane with $k^- = |k^{-}|e^{i\theta_{k}}$ (Fig.~\ref{fig:resonance}), we set the sign of the effective range to be negative. In this case, $k^{-}$ ($k^{+}$) corresponds to resonances (anti-resonances) with positive (negative) real part.
Furthermore, geometric constraint to obtain resonances in Fig.~\ref{fig:resonance} indicates that the magnitude of the effective range $|r_{e}|$ should be larger than the magnitude of the scattering length $|a_{0}|$:
\begin{align}
\frac{1}{|r_{e}|}\sqrt{\frac{2r_{e}}{a_{0}}-1} \geq \frac{1}{|r_{e}|}, \quad \Rightarrow \quad \frac{r_{e}}{a_{0}} \geq 1,\quad \Rightarrow \quad |a_{0}|\leq |r_{e}|,
\label{eq:re}
\end{align}
where we use $r_{e}<0$. This property has an important implication for near-threshold resonances. When a bound state approaches the threshold, the scattering length $|a_{0}|$ becomes much larger than other length scales including $|r_{e}|$, leading to the low-energy universality of shallow bound states~\cite{Braaten:2004rn,Naidon:2016dpf}. When a resonance approaches the threshold with keeping $0 < -\theta_{k} \leq \pi/4 $, one can show from Eq.~\eqref{eq:a-ERE} that $|a_{0}|$ diverges. However, Eq.~\eqref{eq:re} indicates that $|r_{e}|$ also diverges with $|a_{0}|\to \infty$, and the scattering length is not a unique length scale. Therefore, the low-energy universality does not apply to resonances, and near-threshold resonances do not necessarily need to be molecular dominant in contrast to bound states. 

For more quantitative investigation of the internal structure of resonances, we calculate the compositeness. In the ERE, the compositeness $X$ is obtained as~\cite{Hyodo:2013iga}
\begin{align}
X&=\sqrt{\frac{1}{1-\frac{2 r_{e}}{a_{0}}}} .
\label{eq:wbr}
\end{align}
By substituting Eqs.~\eqref{eq:a-ERE} and \eqref{eq:r-ERE}, the compositeness $X$ can be rewritten with the eigenmomenta as
\begin{align}
X &= -\frac{k^{-}+k^{+}}{k^{-}-k^{+}}.
\end{align}
We then use the relation between the resonance and anti-resonance $k^{+}=-k^{-*}$ to obtain
\begin{align}
X&=-i\frac{{\rm Im}\ k^{-}}{{\rm Re}\ k^{-}} = -i\tan \theta_{k}=-i\tan\left(\frac{1}{2}\theta_{E}\right),
\label{eq:X-ERE-theta}
\end{align}
where $\theta_{E}$ stands for the argument of the eigenenergy $E=|E|e^{i\theta_{E}}$. In this way, the compositeness of unstable resonances takes a complex value as mentioned in the introduction. In particular, the compositeness of resonances described by the ERE is pure imaginary~\cite{Hyodo:2013iga}.

\begin{figure}[t]
\centering
\includegraphics[width=0.75\textwidth]{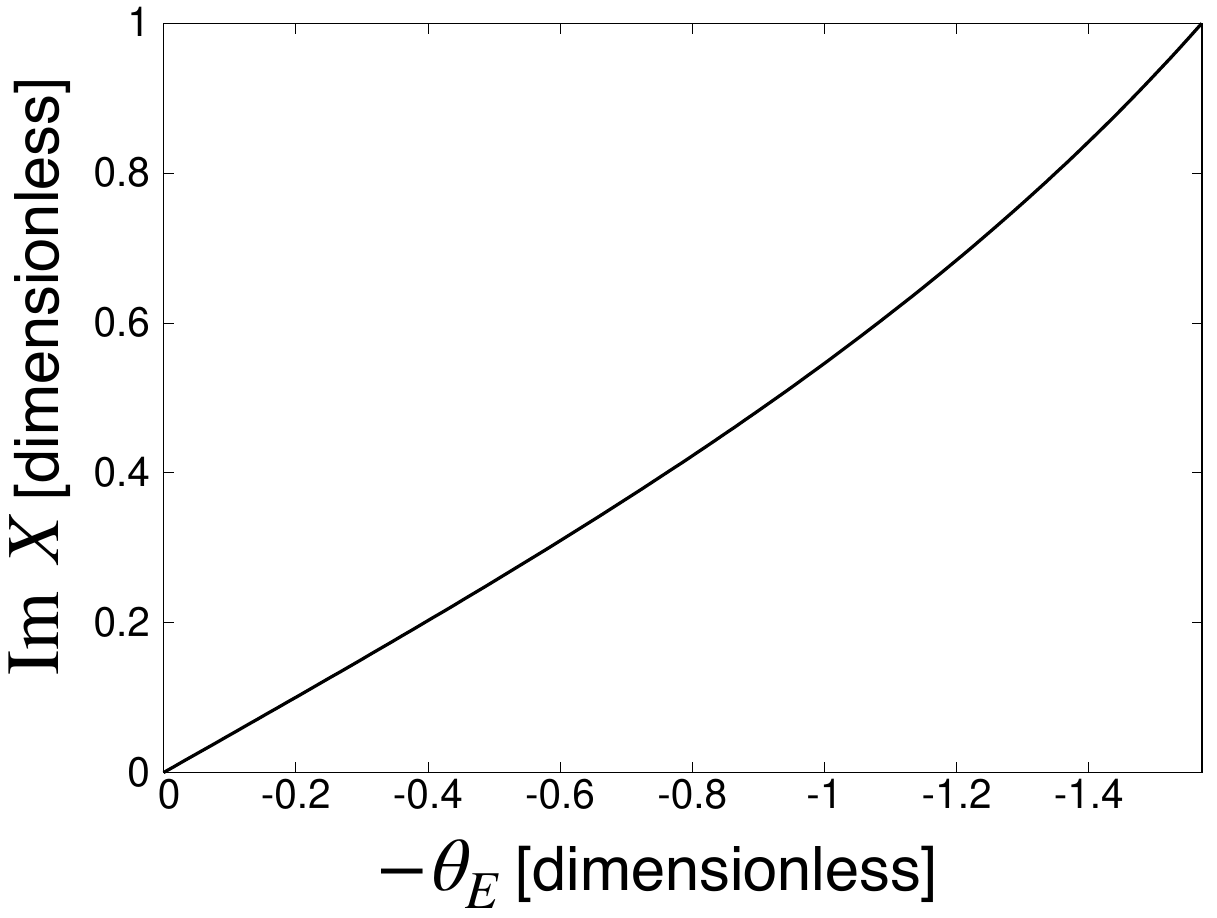}
\caption{Imaginary part of compositeness of resonances in the ERE as a function of the argument of the eigenenergy $-\theta_{E}$. }
\label{fig:complex-X}
\end{figure}

In Fig.~\ref{fig:complex-X}, we show the imaginary part of $X$ of resonances as a function of $-\theta_{E}$ in the forth quadrant in the complex energy plane ($0< -\theta_{E}\leq \pi/2$). We see that Im $X$ is exactly zero on the real axis $\theta_{E} = 0$, and gradually increases with $-\theta_{E}$. Finally, Im $X$ becomes unity on the imaginary axis $-\theta_{E} = \pi/2$. 
 
To analyze the internal structure of resonances, we apply our interpretation scheme in Sec.~\ref{sec:interpretation} to complex $X$ in Eq.~\eqref{eq:X-ERE-theta}. To perform quantitative discussion, we need to determine the explicit value of the parameter $\alpha$. In this study, we regard only resonances with small decay width as physical states. Specifically, we accept resonances whose decay width is smaller than the exciation energy. With this criterion, we determine
\begin{align}
\alpha = \alpha_{0} \equiv \frac{\sqrt{5}-1+\sqrt{10-4\sqrt{5}}}{2} \approx 1.1318, 
\label{eq:alpha}
\end{align}
which excludes resonances with larger decay width $\Gamma$ than the real part of the eigenenergy Re $E$:
\begin{align}
{\rm Re\ }E < \Gamma \quad \Rightarrow \quad |\theta_{E}|> |\arctan(-1/2)|,
\end{align}
as non-interpretable states with negative $\mathcal{X}$. For more details of the determination of $\alpha_{0}$, see Ref.~\cite{Kinugawa:2024kwb}. 

\begin{figure}[t]
\centering
\includegraphics[width=0.75\textwidth]{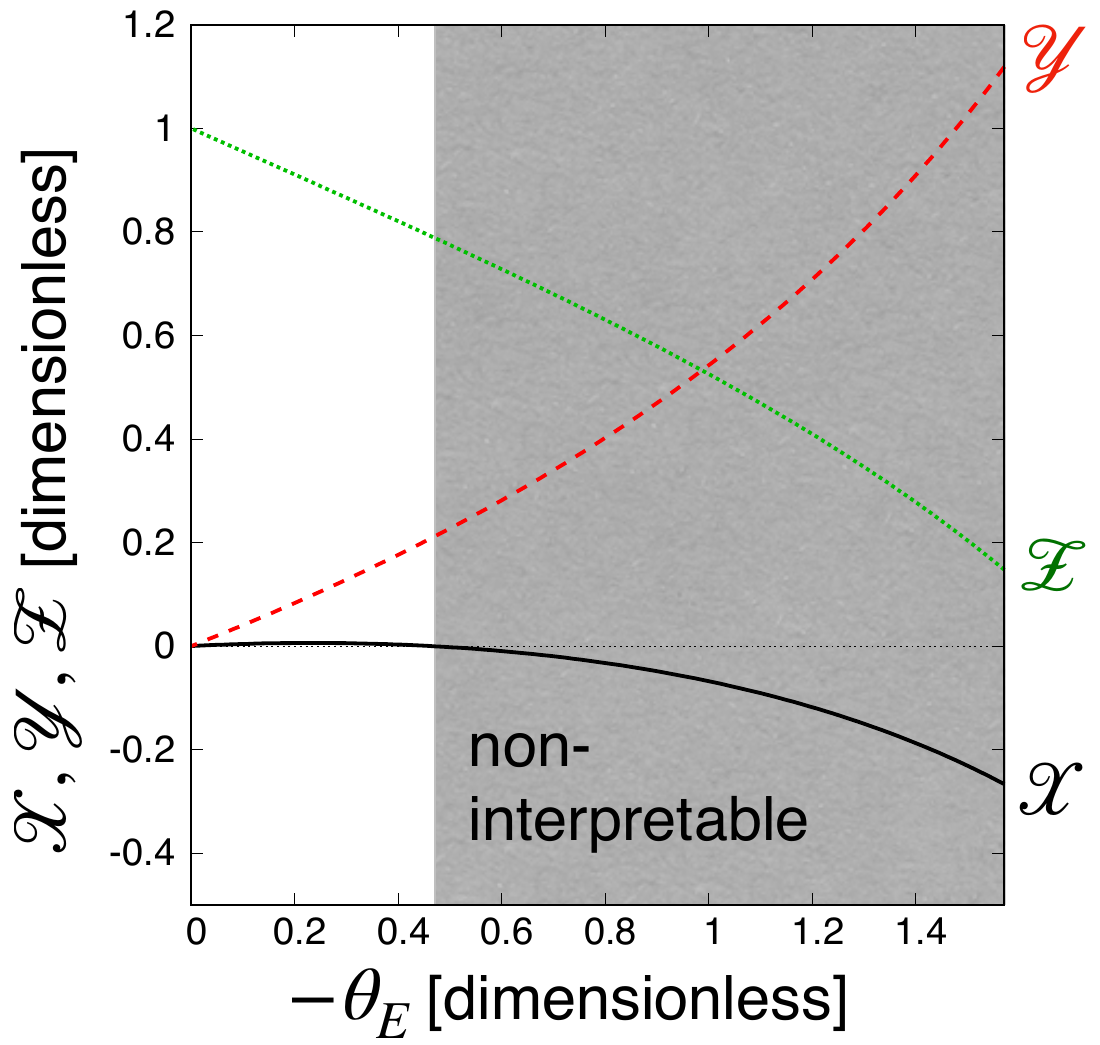}
\caption{The probabilities $\mathcal{X,Y,Z}$ as functions of the argument of the complex energy plane in $0\leq -\theta_{E}  \leq \pi/2$ region with $\alpha_{0}$. The solid line stands for $\mathcal{X}$, dashed for $\mathcal{Y}$, and dotted for $\mathcal{Z}$. The shaded region represents non-interpretable region where the state has a large decay width.}
\label{fig:XYZ}
\end{figure}

With $\alpha_{0}$, we calculate $\mathcal{X,Y,Z}$ from the complex compositeness $X$ in the ERE~\eqref{eq:X-ERE-theta} by varying the argument of the eigenenergy $-\theta_{E}$ in the resonance region $0< -\theta_{E}\leq \pi/2$. In Fig.~\ref{fig:XYZ}, we show $\mathcal{X,Y,Z}$ as functions of $-\theta_{E}$ by the solid, dashed, and dotted lines, respectively. By the definition of $\alpha_{0}$, resonances with a large decay width in the $|\theta_{E}|> |\arctan(-1/2)|$ region are non-interpretable with negative $\mathcal{X}$ (shaded region). In the interpretable region ($|\theta_{E}|< |\arctan(-1/2)|$), all resonances are $\mathcal{Z}$ dominant (non-molecule dominant) with $\mathcal{Z} \gtrsim 0.8$. On the real axis $-\theta_{E} = 0$, the probabilities $\mathcal{X}$ and $\mathcal{Y}$ are exactly zero, and $\mathcal{Z} = 1$. This can be seen analytically by substituting $X=i$ and $Z=1-i$ into Eqs.~\eqref{eq:calX}, \eqref{eq:calY}, and \eqref{eq:calZ}. Intuitively, the pole on the real axis without the decay width has no coupling to the scattering states, and hence the fraction of the scattering components vanishes. When the pole moves away from the real axis, $\mathcal{X}$ and $\mathcal{Y}$ become finite and $\mathcal{Z}$ decreases from unity. However, in the whole interpretable region, $\mathcal{Z}$ is always the largest component, and therefore narrow resonances are non-molecular dominant. Quantitatively, the elementarity of resonances satisfy $\mathcal{Z} \gtrsim 0.8$ in the interpretable region. This non-molecular nature of resonances is sharply contrasted from shallow bound states with $Z \sim 0$, whose molecular nature arises as a consequence of the low-energy universality.

\begin{figure}[t]
\centering
\includegraphics[width=0.75\textwidth]{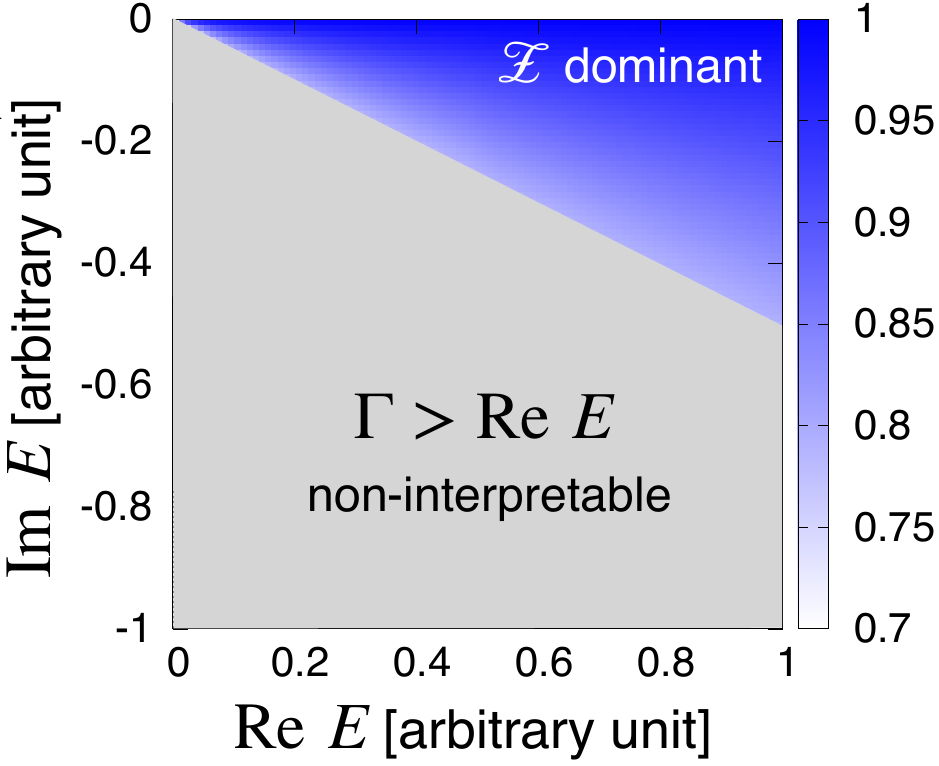}
\caption{Classification of near-threshold resonances in the complex eigenenergy plane.}
\label{fig:Z-non}
\end{figure}

To illustrate this feature, we plot the elementarity $\mathcal{Z}$ in the complex $E$ plane in Fig.~\ref{fig:Z-non}. It is clear that the near-threshold narrow resonances ($\Gamma<{\rm Re}\ E$) are non-molecular dominant, and broad resonances  ($\Gamma>{\rm Re}\ E$) are not interpretable in this scheme. This result is consistent with the fact that the low-energy universality does not hold for resonances, where a larger effective range $|r_{e}|>|a_{0}|$ is required as shown in Eq.~\eqref{eq:re}. However, by shifting the perspective, we find that near-threshold resonances exhibit a kind of universality where the non-molecule dominant nature is determined by both of the scattering length $a_{0}$ and effective range $r_{e}$.

\section{Summary}
\label{sec:summary}

In this work, we have discussed two issues to clarify the molecular nature of near-threshold resonances. In the first part of this study, we propose a new interpretation scheme for unstable resonances with complex compositeness. To take into account the ambiguities of the identification of resonances, we introduce a new probability for ``uncertain identification'' $\mathcal{Y}$ in addition to the compositeness $\mathcal{X}$ and elementarity $\mathcal{Z}$. Furthermore, this scheme allows negative $\mathcal{X}$ or $\mathcal{Z}$ and we utilize this property to define ``non-interpretable'' states so that unphysical resonances with a large decay width are automatically excluded from the interpretation.

As an application of our interpretation scheme, we discuss the internal structure of near-threshold resonances using the effective range expansion. From the relation between the scattering length $a_{0}$ and effective range $r_{e}$, we show that not only $|a_{0}|$ but also $|r_{e}|$ diverges when a resonance approaches the threshold. This is different from the bound state case where only $|a_{0}|$ diverges and low-energy universality governs the molecular nature of the near-threshold bound states. With the quantitative analysis based on our interpretation scheme with $\mathcal{X,Y,Z}$, we find that the near-threshold resonances are generally non-molecular dominant with $\mathcal{Z} \gtrsim 0.8$. Given that the near-threshold shallow bound states are generally molecular dominant, we conclude that the nature of near-threshold state crucially depends on whether the state exists below or above the threshold. 

\bibliographystyle{elsarticle-num.bst}
\bibliography{refs.bib}

\end{document}